\documentclass[twocolumn,english]{revtex4-1}
\usepackage[T1]{fontenc}
\usepackage[latin9]{inputenc}
\usepackage{amsmath}
\usepackage{amssymb}
\usepackage{graphicx}
\usepackage{babel}
\usepackage[pdftex]{hyperref}
\begin{document}
\title{Heat flow and noncommutative quantum mechanics in phase-space}
\author{Jonas F. G. Santos}
\affiliation{Centro de Ciências Naturais e Humanas, Universidade Federal do ABC,
Avenida dos Estados 5001, 09210-580 Santo André, São Paulo, Brazil}
\email{jonas.floriano@ufabc.edu.br, jonasfgs18@gmail.com}

\begin{abstract}
The complete understanding of thermodynamic processes in quantum scales is paramount to
develop theoretical models encompassing a broad class of phenomena as well as to design new
technological devices in which quantum aspects can be useful in areas as quantum information and
quantum computation. Among several quantum effects, the phase-space noncommutativity, which
arises due to a deformed Heisenberg-Weyl algebra, is of fundamental relevance in quantum systems
where quantum signatures and high energy physics play important roles. In low energy physics,
however, it may be relevant to address how a quantum deformed algebra could influence some
general thermodynamic protocols, employing the well-know noncommutative quantum mechanics in
phase-space. In this work, we investigate the heat flow of two interacting quantum systems in the
perspective of noncommutativity phase-space effects and show that by controlling the new constants
introduced in the quantum theory the heat flow from the hot to the cold system may be enhanced,
thus decreasing the time required to reach thermal equilibrium. We also give a brief discussion on
the robustness of the second law of thermodynamics in the context of noncommutative quantum
mechanics.
\end{abstract}
\maketitle

\section{Introduction}

Quantum thermodynamics comprises the energy conversion in a scale
where quantum effects may be useful to improve some specific protocol.
It is expected that advances in quantum thermodynamics will be useful
in many fundamental aspects as well as in a vast range of technological
applications, such as quantum information and quantum communication
\citep{Adesso,Wang}, quantum cryptography \citep{Adesso02,Yang2019,Dowling2015},
quantum computation \citep{Chuangbook,Googlepaper2019,Wright2019},
and in the development of different models of quantum heat machines
\citep{Camati2019,Kosloff2019,Paternostro2019,key-9,Klatzow2019,Santos2018,Camati2020}.
In continuous variable systems, a considerable number of platforms
have been suitable for testing the laws of thermodynamics in the quantum
limit, such as quantum optics \citep{Reiter,Haroche2001,Walls}, optomechanical
devices \citep{Zhang2014,Kampel2013,Clerk2013}, and trapped ions
\citep{Sage2019-1}. Furthermore, supporting the thermal interaction
between two general systems, there is a basic statement claiming that
for initially uncorrelated systems heat naturally flows from the hotter
to the colder system, well know how Clausius statement \citep{Clausius}.
Another particularly interesting scenario in which the heat flow has
been addressed consist in a chain of coupled harmonic oscillators
where the first and last are in contact with thermal reservoirs in
distinct temperatures \citep{Oliveira2017,Casetti2018,Dantan2015},
in cases of several thermal reservoirs in linear quantum lattices
\citep{Landi2019}, and in the presence of time-dependent periodic
drivings \citep{Fiore2019}.

Among relevant quantum effects that have been largely investigated
in quantum thermodynamics, such as entanglement \citep{Horodecki2009},
coherence \citep{Plenio2017}, non-Markovian behavior \citep{Breuer2016}
and general quantum correlations \citep{Modi2012}, from a theoretical
point of view stimulating questions arise when features encompassing
the context of noncommutative phase-space extension of the quantum
mechanics are considered. These questions were firstly addressed in
the configuration space by Synder \citep{Snyder} as a propose to
avoid divergences in the quantum field theory. More recently, with
a deep consensus that in the Planck scale $(\ell_{P}=\sqrt{\hbar G/c^{3}}\sim10^{-35}cm)$
the notion of space-time has to be significantly modified \citep{ref01,ref02},
in order to contemplate general noncommutativity at high energy scales,
a large number of works dealing with noncommutative signatures in
different scenarios has been reported. In what concerns the conventionally
known as noncommutative quantum mechanics (NCQM), there have been
many studies dedicated to investigate possible effects and signatures
of noncommutativity, for instance, in $2D$-harmonic oscillators \citep{Bernardini01,Vergara},
the gravitational quantum well \citep{Orfeu01,Banerjee,Gnatenko2019A, Lawson2017},
in relativistic dispersion relations \citep{Leal01}, and exploring
different aspects of quantum information \citep{Jonas2016,Leal2018,Leal2019,Paris2016}.
The influence of noncommutative quantum mechanics has been also analyzed
in $\mathcal{PT}$-symmetric Hamiltonians \citep{Giri2009,Santos2019A, Dey2012}.
Although at present the ability to experimentally access noncommutative
signatures has not been reached, some theoretical advances have arisen,
e.g. in quantum optics \citep{Dey2017} and optomechanical devices
\citep{Faizal2018}.

Due to the extensive domain of quantum thermodynamics \citep{Anders2016,Kosloff00,Deffner00},
an interesting fundamental question is how noncommutativity could
modify thermodynamics protocols in quantum scales. Motivated by the
same issue, noncommutativity has been addressed in some models of
quantum heat machines \citep{Jonas02,Pandit2019,Chattopadhyay2019}
as well as in dissipative dynamics of Brownian particles \citep{Dias2009,Almeida2017}
and Gaussian states \citep{Santos2019}. In this work we address the
question of how noncommutativity in phase-space could impact the heat
flow between two interacting systems with different temperatures.
With this purpose, we provide a simple recipe to obtain thermal states
of harmonic oscillators that evolve under the action of noncommutative
parameters. Then, evolving these interacting thermal states unitarily,
we show that the noncommutative effects may be used to enhance the
heat flow from the hot to the cold system, i.e. decreasing the time
to the two systems reach thermal equilibrium. On theoretical aspects,
this work is supported by recent studies concerning quantum effects
in the heat flow between interacting systems, for instance, in Ref.
\citep{Partovi2008,Rudolph2012}, and by arguments that quantum signatures
due to additional noncommutativity are universal \citep{Das2008}.
On the other hand, the experimental implementation of the reversion
in the heat flow of two qubits \citep{Serra2019} inspires for searching
new quantum signatures in quantum thermodynamic processes. Moreover,
the use of analog gravity models to simulate general uncertainty has
been reported in Ref. \citep{Conti2019}.

The work is organized as follows. Section \ref{sec:Theoretical-framework}
is dedicated to introduce the main properties of noncommutativity
in phase-space as well as Gaussian states and heat flow. In section
\ref{sec:Enhancing-the-heat} we discuss formal aspects in considering
noncommutative effects during the heat exchange process between two
interacting systems. In the following, we study an example of two
interacting thermal states of harmonic oscillators discussing the
local internal energies and introducing a heating power for the colder
system. We conclude and draw final remarks in section \ref{sec:Conclusions}.

\section{Theoretical framework\label{sec:Theoretical-framework}}

In this section we review the basic theoretical properties of noncommutative
quantum mechanics in phase-space and some important tools of quantum
thermodynamics and quantum information that will be useful in the
following.

\subsection{Noncommutative quantum mechanics in phase-space}

The noncommutativity of the phase-space is based on the deformed Heisenberg-Weyl
algebra \citep{Rojas2001,Bastos2008,Gouba2010,Gouba2016} which can
be represented by the commutation relations,

\begin{equation}
\left[q_{i},q_{j}\right]=i\theta_{ij},\,\left[q_{i},p_{j}\right]=i\hbar\delta_{ij},\,\left[p_{i},p_{j}\right]=i\eta_{ij},\label{ncrelation}
\end{equation}
with $i,j=1,...,d$, $\theta_{ij}$ and $\eta_{ij}$ are invertible
antisymmetric real constant $(d\times d)$ matrices, and one can define
the matrix $\Sigma_{ij}=\delta_{ij}+\theta_{ik}\eta_{kj}/\hbar^{2}$,
which is also invertible if $\theta_{ik}\eta_{kj}\neq-\hbar^{2}\delta_{ij}$.
Writing $\theta_{ij}=\theta\epsilon_{ij}$ and $\eta_{ij}=\eta\epsilon_{ij}$,
with $\epsilon_{ii}=0$, $\epsilon_{ij}=-\epsilon_{ji}$, one can
interpret $\theta$ and $\eta$ as being new constants in the quantum
theory, which have been extensively studied recently \citep{Bernardini01,Jonas02,Saha,Bastos,Bastos02,Andreas,Liang2019,Bastos2008-1}. Besides, $\theta$ and $\eta$ are assumed to be positive quantities.
Given a quantum system described by the Hilbert space of the NCQM,
it is possible to represent it in the context of the standard quantum
mechanics (SQM), which is governed by the well-know commutation relations,

\begin{equation}
\left[Q_{i},P_{j}\right]=0,\,\left[Q_{i},P_{j}\right]=i\hbar\delta_{ij},\,\left[P_{i},P_{j}\right]=0.
\end{equation}

This is performed through the Seiberg-Witten (SW) map \cite{Orfeu02, Santos2015, Gamboa}, given by,

\begin{equation}
q_{i}=\nu Q_{i}-(\theta/2\nu\hbar)\epsilon_{ij}P_{j},\quad p_{i}=\mu P_{i}+(\eta/2\mu\hbar)\epsilon_{ij}Q_{j},\label{SW}
\end{equation}
in which $\nu$ and $\mu$ are arbitrary parameters fulfilling the
condition $\theta\eta=4\hbar^{2}\mu\nu(1-\mu\nu)$. Thus, for a general
quantum system described by the Hamiltonian $H^{nc}(q_{i},p_{i})$
in the NC phase-space, the action of the SW map can be summarized
as follows,

\begin{align}
H^{nc}(q_{i},p_{i}) & \longrightarrow\mathcal{H}(Q_{i},P_{i})\nonumber \\
 & =H(Q_{i},P_{i})+f(\theta,\eta)V(Q_{i},P_{i}),\label{generalrole}
\end{align}
where $H(Q_{i},P_{i})$ has the same Hamiltonian structure as $H^{nc}(q_{i},p_{i})$,
$f(\theta,\eta)$ is a function of the noncommutative parameters,
and $V(Q_{i},P_{i})$ is in general an interaction Hamiltonian term.

\subsection{Thermal states and heat flow}

Thermal states have special relevance in quantum thermodynamics, for
instance, they are useful in representing asymptotic states of quantum
systems in contact with thermal reservoirs and to build several models
of quantum thermal machines. Moreover, they are a particular set of
a larger one, well know as Gaussian states which, in its turn, are
completely characterized by the first statistical moments and covariance
matrix \citep{Adesso,Wang,Serafini}. For two-mode Gaussian states,
defining a vector $\vec{R}(Q_{1},P_{1},Q_{2},P_{2})$ to group the
coordinates of a two-dimensional system, the first moments are defined
as the vector $\vec{d}=(\langle Q_{1}\rangle_{\rho},\langle P_{1}\rangle_{\rho},\langle Q_{2}\rangle_{\rho},\langle P_{2}\rangle_{\rho})$.
The covariance matrix (CM) is the set of all second statistical moments,
given by $\sigma=\sigma_{11}\oplus\sigma_{22}$ for a two-mode Gaussian
state, with,

\begin{equation}
\sigma_{ii}=\left(\begin{array}{cc}
\sigma_{Q_{i}Q_{i}} & \sigma_{P_{i}Q_{i}}\\
\sigma_{Q_{i}P_{i}} & \sigma_{P_{i}P_{i}}
\end{array}\right),\label{CM}
\end{equation}
where $\sigma_{\alpha\beta}=\langle\alpha\beta+\beta\alpha\rangle_{\rho}-2\langle\alpha\rangle_{\rho}\langle\beta\rangle_{\rho}$.
For physical two-mode Gaussian states the CM satisfies the relation
$\sigma+i\Omega\geq0$, with, 

\begin{equation}
\Omega=\left(\begin{array}{cc}
0 & 1\\
-1 & 0
\end{array}\right)^{\oplus2},
\end{equation}
 such that $\left[\vec{R}_{i},\vec{R}_{j}\right]=i\Omega_{ij}$ \citep{Simon}.
In special, for Gaussian states the Wigner function assumes a gently
form given by \citep{Adesso},

\begin{equation}
W_{G}(\vec{R})=\frac{exp\left[-(1/2)(\vec{R}-\vec{d})\sigma^{-1}(\vec{R}-\vec{d})\right]}{(2\pi)^{2n}\sqrt{Det[\sigma]}}.\label{wigner}
\end{equation}
A one-mode thermal state can be represented by \citep{Adesso,Wang,Serafini},

\begin{equation}
\rho^{th}(\bar{n})=\sum_{n=0}^{\infty}\frac{\bar{n}^{m}}{(\bar{n}+1)^{n+1}}|n\rangle\langle n|,\label{therm}
\end{equation}
with $\bar{n}=\left\{ \text{exp}\left[\hbar\omega/k_{B}T\right]-1\right\} ^{-1}$
the mean number of photons in the bosonic mode, $k_{B}$ is the Boltzmann
constant, $T$ is the associated temperature, and $\{|m\rangle\}$
is the Fock basis. It is direct to note that number $\bar{n}$ is
proportional to the temperature of a thermal state. The first moments
and covariance matrix of a one-mode thermal state are $\vec{R}=(0,0)$
and $\sigma=(2\bar{n}+1)\mathbb{I}_{2\times2}$, respectively \citep{Adesso,Wang,Serafini}.

\begin{figure}
\includegraphics[scale=0.7]{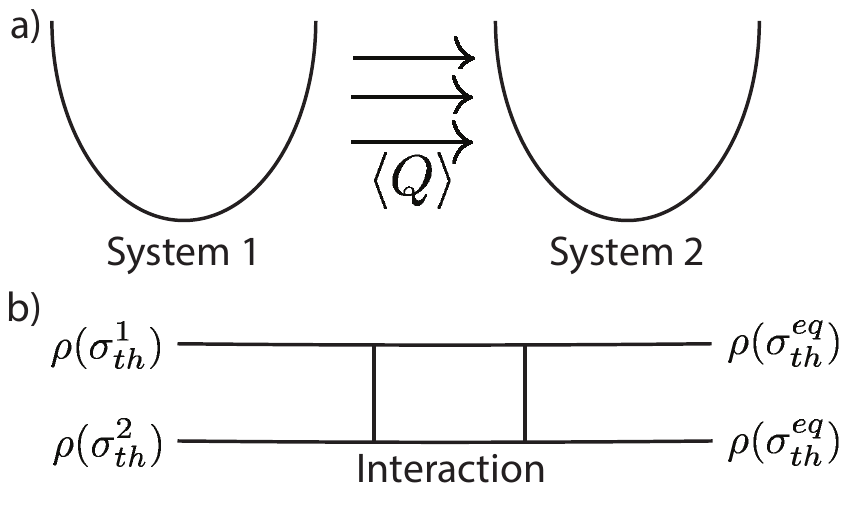}

\caption{a) Illustration of the heat exchange process in which heat flows from
the system 1 to the system 2. b) A simple circuit depicting the interaction
between the two systems, initially prepared in local thermal states
and achieving thermal equilibrium at the end of the protocol.}

\label{illustration}
\end{figure}

In quantum thermodynamics, given a quantum system with time-independent
Hamiltonian $H$ and state $\rho$, the associated energy is obtained
by $U(\rho(t))=\text{Tr}[H\rho(t)]$. For the particular case of Gaussian
states with zero first moments, the energy is simply the trace of
the covariance matrix, i.e. $U(\rho(t))=\text{\ensuremath{(\hbar\omega/4)}Tr}[\sigma(t)]$
\citep{Adesso} (see \ref{sec:Appendix-1}). Considering two quantum
systems, $1$ and $2$, under the action of a heat exchange protocol
during the time comprised between $[0,\tau]$, the heat exchanged
by the system $1$ is given by $\langle Q\rangle_{1}=\text{\ensuremath{(\hbar\omega/4)(}Tr}[\sigma(\tau)]-\text{Tr}[\sigma(0)])$
and, as expected, $\langle Q\rangle_{2}=-\langle Q\rangle_{1}$, since
the we are assuming a unitary evolution, as illustrated in Fig. \ref{illustration}.

\section{Enhancing the heat flow with nc effects\label{sec:Enhancing-the-heat}}

\subsection*{General formalism}

Consider two interacting systems in which the time-independent Hamiltonian
$H(q_{i},p_{i})$ is governed by the commutation relations in (\ref{ncrelation}).
After implementing the Seiberg-Witten map, the new Hamiltonian is
given by $\mathcal{H}(Q_{i},P_{i})$. Assuming that we have initially
uncorrelated local thermal states for each system such that, $\rho_{i}^{th}(0)=\text{exp}(-\beta_{i}\hbar\omega_{i})/\text{Tr}[\exp(-\beta_{i}\hbar\omega_{i})]$,
after the time evolution the final states are,

\[
\rho_{i}^{th}(t)=U_{t,0}^{\theta,\eta}\rho_{i}^{th}(0)\left(U_{t,0}^{\theta,\eta}\right)^{\dagger},
\]
with the time-evolution operator given by $U_{t,0}^{\theta,\eta}=\exp[-i\mathcal{H}t/\hbar]$.
It is worth to mention that, once the structure of $\mathcal{H}(Q_{i},P_{i})$
is given by Eq. (\ref{generalrole}), the time evolution of the composite
system generates correlations between the parts. The internal energy
of each system in time $t$ is $E_{i}(t)=Tr[\rho_{i}^{th}(t)\mathcal{H}]$
which allows to write the heat exchanged during the time evolution,
$\langle Q_{i}\rangle=E_{i}(t)-E_{i}(0)$. In the absence of any noncommutative
effects, $\theta=\eta=0$, we recover the traditional form of the
heat well know in the standard quantum mechanics. However, the existence
of any noncommutative effects will be captured in the heat exchanged
by the two systems. A more detailed information about the time-evolution
operator with NC effects is provided in \ref{sec:Appendix}.

\subsection*{Two-interacting oscillators}

Here we consider an example of two interacting quantum systems in
order to show that the noncommutativity of the phase-space may be
used to enhance the heat flow and decrease the time to reach the thermal
equilibrium. Consider two-coupled harmonic oscillators described in
the NCQM (commutation relations in Eqs (\ref{ncrelation})) , in which the Hamiltonian reads,

\begin{equation}
H^{nc}(q_{i},p_{i})=\sum_{i=1}^{2}\left(\frac{p_{i}^{2}}{2m}+\frac{m\Omega^{2}}{2}q_{i}^{2}\right)+\frac{\omega_{B}}{2}\sum_{i,j=1}^{2}\epsilon_{ij}p_{i}q_{j,}\label{NCH}
\end{equation}
with $\Omega^{2}=\omega^{2}+\omega_{B}^{2}/4$, in which $m$ and
$\omega$ are the mass and frequency of the oscillators, respectively,
and $\omega_{B}$ is the coupling frequency. This type of interaction
corresponds, for instance, to a uniform magnetic field applied on
the orthogonal direction, being a good approximation of a confinement
implemented in semiconductor quantum dots \citep{Jacak1998}. Once
we transform the Hamiltonian in Eq. (\ref{NCH}) to the standard quantum
mechanics, we obtain,

\begin{equation}
\mathcal{H}(Q_{i},P_{i})=\alpha^{2}Q_{i}^{2}+\beta^{2}P_{i}^{2}+\left(\frac{\omega_{B}}{2}+\gamma\right)\sum_{i,j=1}^{2}\epsilon_{ij}P_{i}Q_{j,}\label{SH}
\end{equation}
 with the following definitions,

\begin{eqnarray*}
\alpha^{2} & = & \frac{\nu^{2}m\Omega^{2}}{2}+\frac{\eta^{2}}{8m\mu^{2}\hbar^{2}}+\frac{\nu}{\mu}\frac{\omega_{B}\eta}{4\hbar^{2}},\\
\beta^{2} & = & \frac{\mu^{2}}{2m}+\frac{m\Omega^{2}\theta^{2}}{8\nu^{2}\hbar^{2}}+\frac{\mu}{\nu}\frac{\omega_{B}\theta}{4\hbar},\\
\gamma & = & \frac{\theta}{2\hbar}m\Omega^{2}+\frac{\eta}{2m\hbar}.
\end{eqnarray*}

Despite the relative complexity in expressions for $\alpha^{2}$ and
$\beta^{2}$, we note that the only effect of these constants on the
Hamiltonian is to induce a shift in the position and momentum of the
oscillators. The relevant effect is exclusively due to $\gamma$ which
depends on the NC parameters and may influence the interaction Hamiltonian,

\begin{equation}
V(Q_{i},P_{i})=\left(\frac{\omega_{B}}{2}+\gamma\right)\sum_{i,j=1}^{2}\epsilon_{ij}P_{i}Q_{j}.\label{InteractionHamiltonian}
\end{equation}

It is important to note that the interaction Hamiltonian commutes
with the total Hamiltonian of the two quantum oscillators, $[\alpha^{2}Q_{i}^{2}+\beta^{2}P_{i}^{2},V(Q_{i},P_{i})]=0$,
implying that the heat exchange process that we will employ does not
perform work. This results that the heat exchanged between the oscillators
is constant in time, $\langle Q_{1}\rangle+\langle Q_{2}\rangle=0$,
with the internal energy variation, \textbf{$\Delta E_{i}=\langle Q_{i}\rangle=Tr[\sigma_{i}(\tau)]-Tr[\sigma_{i}(0)]$}.

To model the thermal interaction between the two oscillators we consider
the initial state preparation developed in Ref. \citep{Bernardini01}
and with the Wigner function written as,

\begin{equation}
W_{k,\ell}(Q_{i},P_{i})=\frac{(-1)^{k+\ell}}{\pi^{2}\hbar^{2}}\exp\left[-\xi_{1(2)}^{2}/\hbar\right]L_{k(\ell)}^{(0)}\left[2\xi_{1(2)}^{2}/\hbar\right],\label{Wigner}
\end{equation}
 with,

\begin{align*}
\xi_{1}^{2} & =\left(\frac{\alpha}{\beta}Q_{1}^{2}+\frac{\beta}{\alpha}P_{1}^{2}\right)\cos(\Gamma t)^{2}+\left(\frac{\alpha}{\beta}Q_{2}^{2}+\frac{\beta}{\alpha}P_{2}^{2}\right)\sin(\Gamma t)^{2}\\
 & -\left(\frac{\alpha}{\beta}Q_{1}Q_{2}+\frac{\beta}{\alpha}P_{1}P_{2}\right)\sin(2\Gamma t),\\
\xi_{2}^{2} & =\left(\frac{\alpha}{\beta}Q_{1}^{2}+\frac{\beta}{\alpha}P_{1}^{2}\right)\sin(\Gamma t)^{2}+\left(\frac{\alpha}{\beta}Q_{2}^{2}+\frac{\beta}{\alpha}Q_{2}^{2}\right)\cos(\Gamma t)^{2}\\
 & +\left(\frac{\alpha}{\beta}Q_{1}Q_{2}+\frac{\beta}{\alpha}P_{1}P_{2}\right)\sin(2\Gamma t),
\end{align*}
with $k$ and $\ell$ integer numbers and $\Gamma=\omega_{B}/2+\gamma$.
Note that the state $W_{k,\ell}(q_{i},p_{i})$ is non-stationary unless
$\Gamma=0$, implying no interaction between the oscillators. Then,
the role of the noncommutative parameters, mediated by $\gamma$,
is to affect the interaction between the systems, as evidenced in
Eq. (\ref{InteractionHamiltonian}). Starting from Eq. (\ref{Wigner})
we can generate a set of thermal states simply obtaining the covariance
matrix from the local states. First, the local states of each oscillator
are given by tracing out the other degree of freedom, i.e.,

\begin{align}
W_{k,\ell}^{1}=W_{k,\ell}^{1}(Q_{1},P_{1}) & =\int\int dQ_{2}dP_{2}\,W_{k,\ell}(Q_{i},P_{i}),\nonumber \\
W_{k,\ell}^{2}=W_{k,\ell}^{2}(Q_{2},P_{2}) & =\int\int dQ_{1}dP_{1}\,W_{k,\ell}(Q_{i},P_{i}).\label{partialtrace}
\end{align}

From the local states, we then obtain the covariance matrix, just
as in Eq. (\ref{CM}), and define the following local thermal states
as,

\begin{equation}
\sigma_{th}^{i;k,\ell}(t)=\frac{\hbar\omega}{4}(2\bar{n}+1)\left(\begin{array}{cc}
\langle Q_{i}Q_{i}\rangle_{W_{k,\ell}^{i}} & \langle Q_{i}P_{i}\rangle_{W_{k,\ell}^{i}}\\
\langle P_{i}Q_{i}\rangle_{W_{k,\ell}^{i}} & \langle P_{i}P_{i}\rangle_{W_{k,\ell}^{i}}
\end{array}\right),\label{sigma}
\end{equation}
with $i=1,2$ and $\bar{n}$ different for each oscillator. We highlight
that by construction, we have $\langle Q_{i}P_{i}\rangle_{W_{k,\ell}^{i}}=\langle P_{i}Q_{i}\rangle_{W_{k,\ell}^{i}}=0$
and $\langle Q_{i}Q_{i}\rangle_{W_{k,\ell}^{i}}=\langle P_{i}P_{i}\rangle_{W_{k,\ell}^{i}}$,
besides the fact the first moments are zero, characterizing the thermal
states. These results are valid irrespective the values of $k$ and
$\ell$. It is important to stress that the thermal states generated
by these protocol are strictly dependent on the noncommutative parameters
$\theta$ and $\eta$. The internal energy associated to each system
is given by $E_{i}^{k,\ell}(t)=(\hbar\omega/4)Tr[\sigma_{th}^{i;k,\ell}(t)]$,
resulting that the heat exchanging between the two systems is $\langle Q_{1}^{k,\ell}\rangle=-\langle Q_{2}^{k,\ell}\rangle=E_{1}^{k,\ell}(t)-E_{1}^{k,\ell}(0)$.

In order to consider a pair of thermal states to analyze the NC effects
on the heat exchange process, we choose the quantum numbers in the
state (\ref{Wigner}) to be $(k,\ell)=(0,1)$, and make the notation
simpler by writing $\sigma_{th}^{i;k,\ell}=\sigma_{th}^{i}$. Then,
using the protocol to obtain the covariance matrix given in Eq. (\ref{sigma})
we get,

\begin{align}
\sigma_{th}^{1}(t) & =\frac{\hbar\omega}{2}(2\bar{n}+1)\left[1-\frac{1}{2}\cos\left(\omega_{B}\left(\frac{\gamma}{\omega_{B}}+\frac{1}{2}\right)t\right)\right]\mathbb{I}_{2\times2},\nonumber \\
\sigma_{th}^{2}(t) & =\frac{\hbar\omega}{4}(2\bar{m}+1)\left[2+\cos\left(\omega_{B}\left(\frac{\gamma}{\omega_{B}}+\frac{1}{2}\right)t\right)\right]\mathbb{I}_{2\times2}.\label{Cov_th}
\end{align}

Using these covariance matrices for the generated thermal states,
we directly obtain the internal energy associated to each oscillator,

\begin{align}
E_{1}(t) & =\hbar\omega(2\bar{n}+1)\left[1-\frac{1}{2}\cos\left(\omega_{B}\left(\frac{\gamma}{\omega_{B}}+\frac{1}{2}\right)t\right)\right],\nonumber \\
E_{2}(t) & =\frac{\hbar\omega}{2}(2\bar{m}+1)\left[2+\cos\left(\omega_{B}\left(\frac{\gamma}{\omega_{B}}+\frac{1}{2}\right)t\right)\right].\label{Energy}
\end{align}

\begin{figure}
\includegraphics[scale=0.6]{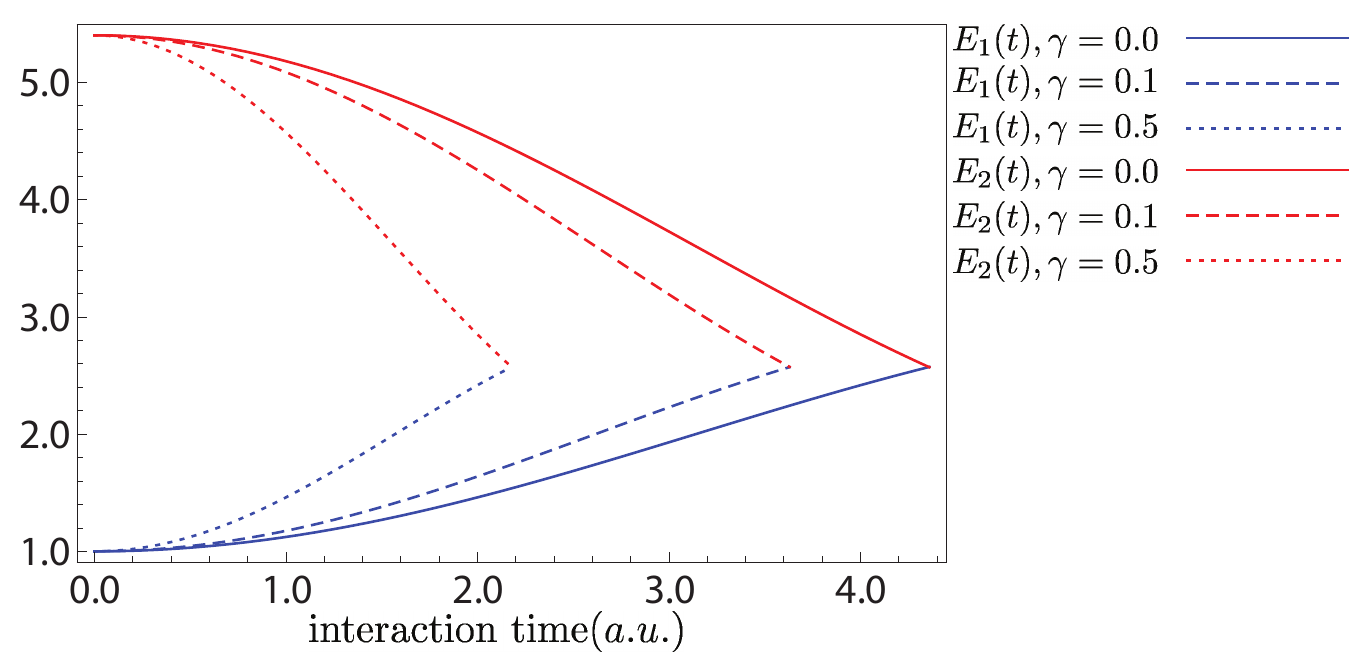}

\caption{Internal energy of the oscillators 1 (red lines) and 2 (blue lines)
as function of the interaction time (arbitrary unities) for different
values of $\gamma$, mediating the NC effects. $\gamma=0$ (solid
lines), $\gamma=0.1$ (dashed lines), and $\gamma=0.5$ (dotted lines).
We have plotted the internal energies for each value of noncommutative
parameter up to the corresponding time required to reach thermal equilibrium.
It was considered $\hbar=m=k_{B}=\omega_{B}=1$, $\omega=4$ and $(\bar{n},\bar{m})=(2,4)$.}

\label{heat_flow}
\end{figure}

\begin{figure}
\includegraphics[scale=0.225]{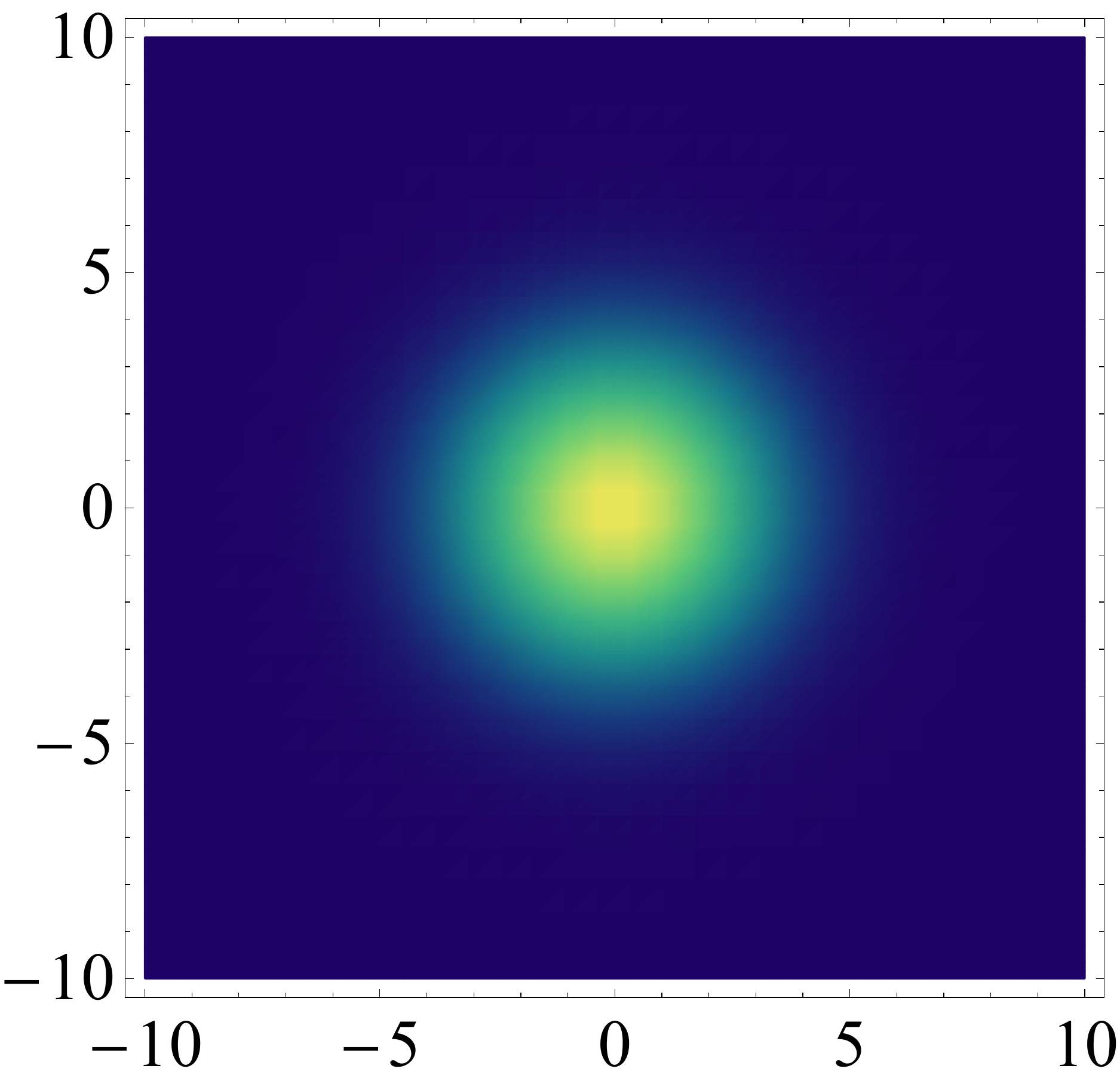}\includegraphics[scale=0.225]{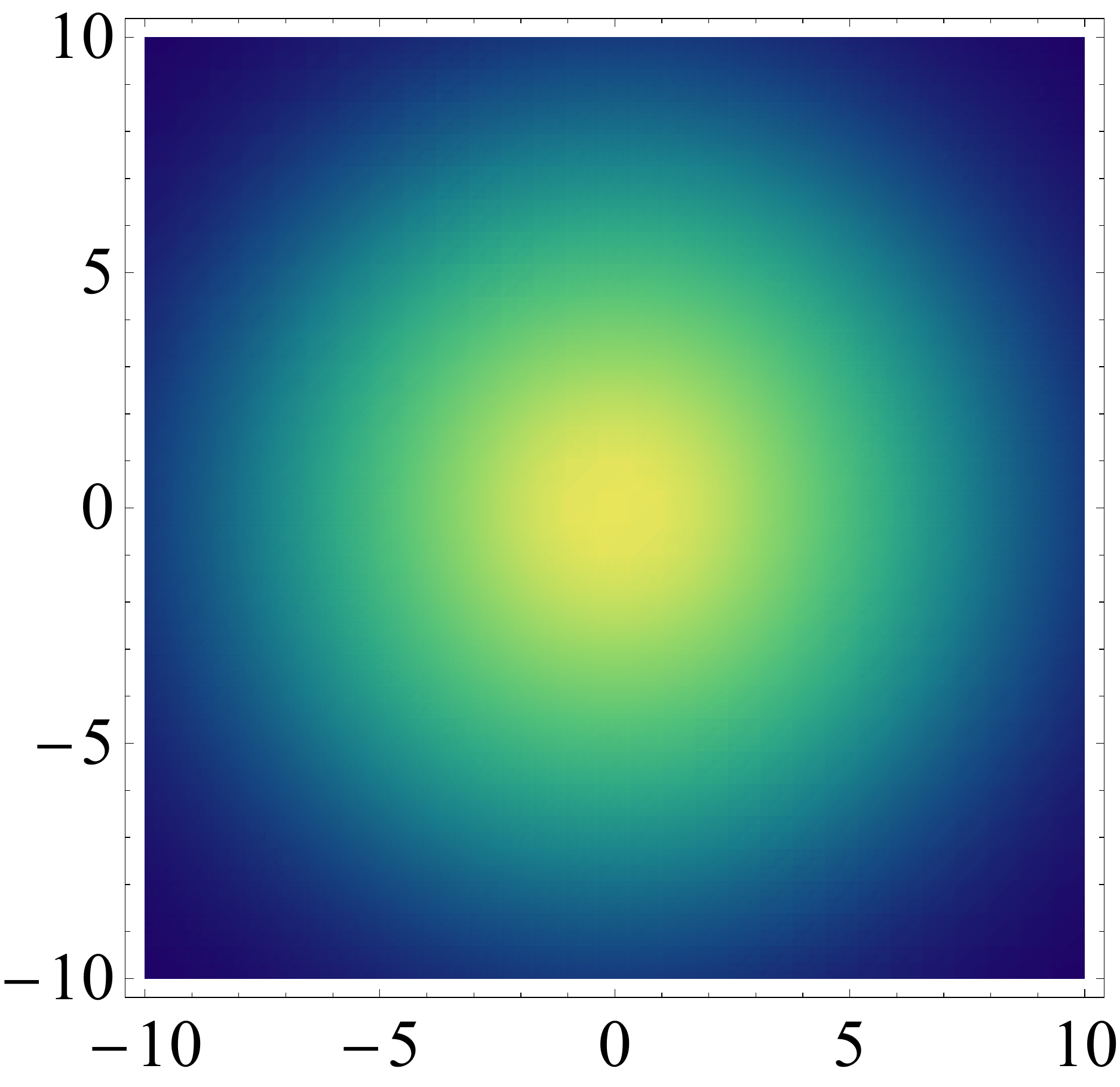}

\includegraphics[scale=0.225]{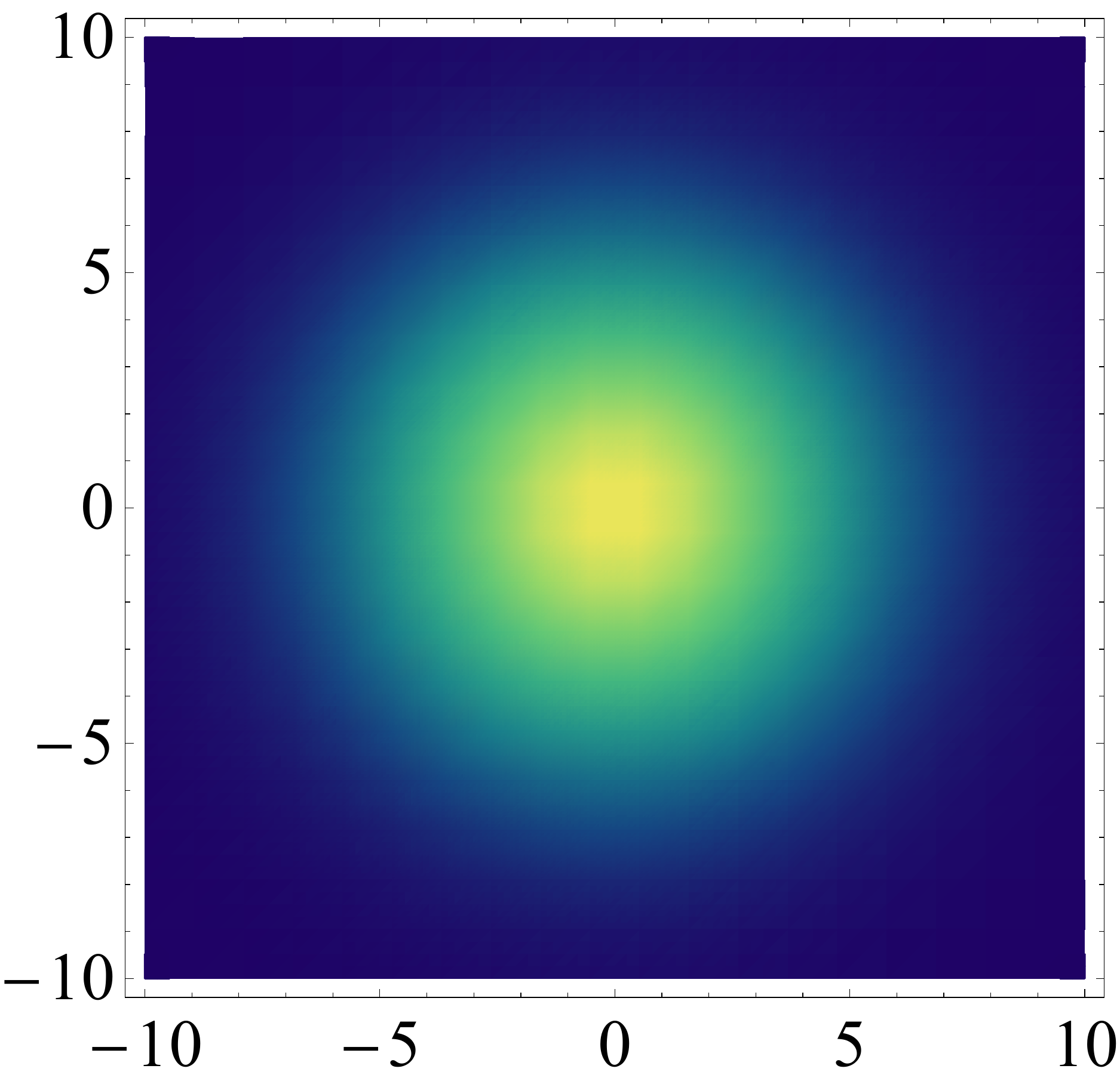}\includegraphics[scale=0.225]{thermal_final}

\label{thermalstates}\caption{Wigner functions (in $(Q_{i},P_{i})$ coordinates) illustrating the
thermal states whose covariance matrices are given by expressions
in Eq. (\ref{Cov_th}). On the top we have the initial thermal states
for $\sigma_{th}^{1}(0)$ (left) and $\sigma_{th}^{2}(0)$ (right),
which do not depend on the NC parameters. The thermal states with
the same internal energy after the heat exchange protocol are depicted
on the bottom. We have considered $\bar{n}=2$ and $\bar{m}=4$, and
a blue-green-yellow color scheme.}
\end{figure}

In Fig. (\ref{heat_flow}) we show the internal energy of the two
oscillators during the heat exchange protocol for different values
of NC parameters. We have chosen $(\bar{n},\bar{m})=(2,4)$, meaning
that the oscillator 2 (red lines) is hotter than the oscillator 1
(blue lines). The case without any NC effects, $\gamma=0$, is represented
by the solid lines, whereas we considered two cases with NC effects,
$\gamma=0.1$ (dashed lines), and $\gamma=0.5$ (dotted lines). It
can be observed that the inclusion of a deformed Heisenberg-Weyl algebra,
see Eq. (\ref{ncrelation}), allows to decrease the time required
to reach thermal equilibrium. Furthermore, in order to illustrate
the initial thermal states given by the covariance matrices in Eq.
(\ref{Cov_th}) as well as the thermal states in the thermal equilibrium,
we plot in Fig. (\ref{thermalstates}) the corresponding Wigner function.
The time to reach thermal equilibrium is found to be $\tau=2\arccos[-2(m-n)/(1+m+n)]/(2\gamma+\omega_{B})$.
A similar result is obtained in the case of $(k,\ell)=(1,0)$.

\begin{figure}
\includegraphics[scale=0.6]{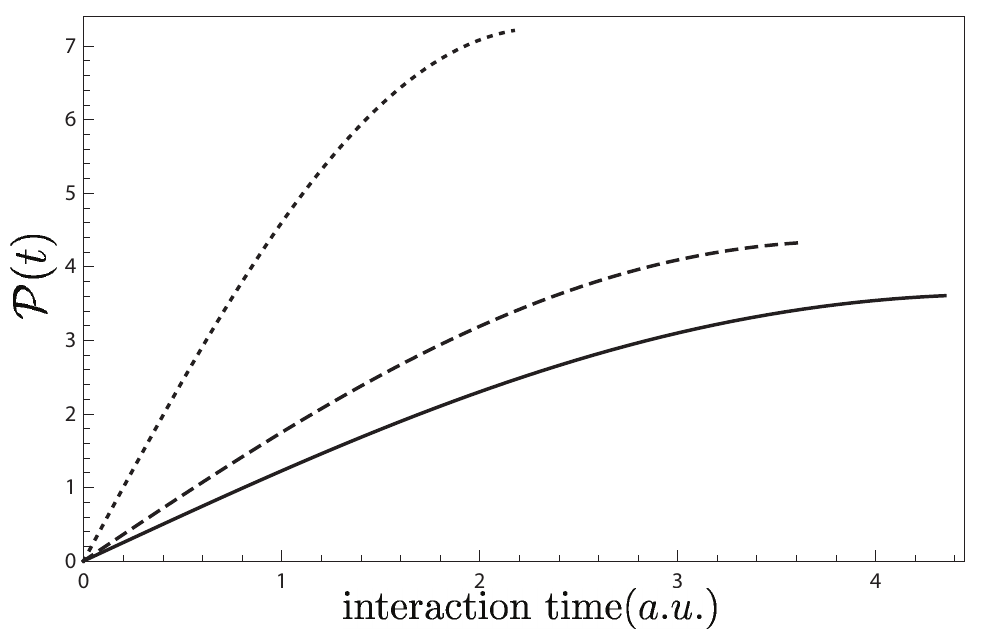}

\caption{Heating power as a function of the interaction time for the same parameters
used in Fig. (\ref{heat_flow}). We have plotted the heating power
for each value of noncommutative parameter up to the corresponding
time required to reach thermal equilibrium.}

\label{heating}
\end{figure}

To explicitly show how noncommutative effects can enhance the heat
exchange process we define a heating power, which is the internal
energy variation of the colder system divided by the time to reach
thermal equilibrium,

\begin{align}
\mathcal{P}(t) & =\frac{E_{1}(t)-E_{1}(0)}{t}.\label{power}
\end{align}
 Figure \ref{heating} presents the heating power as a function of
the interaction time for the same parameters used in Fig. (\ref{heat_flow}).
It is possible to note that the decrease in the thermal equilibrium
time due to noncommutative effects implies in larges values of $\mathcal{P}(t)$.

\subsection*{Second law of thermodynamics and NC effects}

We would like to discuss briefly the possible impacts of the noncommutative
parameters on the second law of thermodynamics. For initially uncorrelated
two thermal states, with temperatures $T_{2}\geq T_{1}$, or $\bar{n}_{2}\geq\bar{n}_{1}$,
a standard way of writing the second law of thermodynamics is given
by \citep{Partovi2008,Rudolph2012},

\[
Q_{1}\left(\frac{1}{k_{B}T_{1}}-\frac{1}{k_{B}T_{2}}\right)\geq0,
\]
meaning that heat flows from the hotter to the colder system. For
the case we have considered, it is possible to note that the inclusion
of new noncommutative relations do not affect the validity of the
standard second law of thermodynamics.

\section{Conclusions\label{sec:Conclusions}}

Quantum thermodynamics has demonstrated its ability in many sectors
of physics in quantum scales, experimental and theoretically. Here
we have assumed a deformed Heisenberg-Weyl algebra and investigated
how the noncommutative parameters $\theta$ and $\eta$ may influence
the heat flow of two coupled harmonic oscillators interacting through
a heat exchange protocol.

Starting from a general preparation of states proposed in Ref. \citep{Bernardini01}
whose dynamics is dictated by NC parameters, we provided a scheme
for obtaining thermal states that are coupled via the interaction
Hamiltonian containing NC effects. Then, since that the interaction
Hamiltonian commutes with the total Hamiltonian of the individual
systems, it is ensured that all the heat flowing out from the system
2 is absorbed by the system 1. The results shows that noncommutative
effects may be employed to enhance the heat flow between the two oscillators,
decreasing the time required to reach thermal equilibrium. We highlight
that since the noncommutative parameters $\theta$ and $\eta$ are
positive quantities, it is not possible to employ it to reverse the
heat flow, implying that the standard second law of thermodynamics
is robust to the inclusion of new noncommutative relations in the
quantum theory. Our results could be used to generate a quantum Otto
refrigerator, with NC effects boosting the performance of the quantum
fridge.

Although at the moment the technological ability does not allow to
experimentally access quantum effects in the Planck scale, our results
can be perfectly simulated using some different platforms, such as
optomechanical and optical devices. We hope that this work can contribute
in this direction, helping to elucidate the role of noncommutative
effects in thermodynamic protocols.

\begin{acknowledgments}
Jonas F. G. Santos acknowledges São Paulo Research Foundation (FAPESP),
Grant No. 2019/04184-5 and Federal University of ABC for support.
\end{acknowledgments}

\section{DATA AVAILABILITY}

Data sharing is not applicable to this article as no new data were created or analyzed in this
study.

\setcounter{section}{0}
\global\long\def\thesection{Appendix \Alph{section}}

\section{Energy and covariance matrix for thermal states\label{sec:Appendix-1}}

Consider a simple quantum harmonic oscillator with frequency $\omega$
and mass $m$ given by,

\[
H(Q,P)=\frac{1}{2m}P^{2}+\frac{m\omega^{2}}{2}Q^{2}.
\]

The mean value of energy is given by $\langle E\rangle_{\rho}=\langle H\rangle_{\rho}$,
with $\rho$ a thermal state. Then,

\[
\langle E\rangle_{\rho}=\frac{1}{2m}\langle P\rangle_{\rho}^{2}+\frac{m\omega^{2}}{2}\langle Q\rangle_{\rho}^{2}=\hbar\omega(\langle a^{\dagger}a\rangle_{\rho}+1/2),
\]
 where $a$ and $a^{\dagger}$ are the annihilation and creation operators,
respectively, with $[a,a^{\dagger}]=1$. From \citep{Adesso} we have
$\langle a^{\dagger}a\rangle_{\rho}=(\text{Tr}[\sigma]-2)/4$, with
$\sigma$ the covariance matrix associated to the thermal state $\rho$,
$\sigma_{ij}=\langle d_{i}d_{j}+d_{j}d_{i}\rangle_{\rho}-2\langle d_{i}\rangle_{\rho}\langle d_{j}\rangle_{\rho}$
and $\vec{d}=(q,p)$. Thus we have,

\[
\langle E\rangle=\frac{\hbar\omega}{4}\text{Tr}[\sigma].
\]

\section{Time evolution of thermal states with NC effects\label{sec:Appendix}}

Here we provide more details about the time evolution of thermal states
with noncommutative effects. Consider that the interaction between
the two systems in the noncommutative phase-space coordinates is represented
by the Hamiltonian $H_{I}(q_{i},p_{i}).$ After the Seiberg-Witten
map, the general interaction reads $g(\theta,\eta)H_{I}(Q_{i},P_{i})+f(\theta,\eta)V(Q_{i},P_{i})$,
with $g(\theta,\eta)$ and $f(\theta,\eta)$ real functions such that
$g(\theta,\eta)=1$ and $f(\theta,\eta)=0$ in the absence of NC effects.
In the interaction picture the time evolution of the local thermal
states is given by,

\begin{align*}
\rho_{i}^{th}(t) & =U_{t,0}^{\theta,\eta}\rho_{i}^{th}(0)\left(U_{t,0}^{\theta,\eta}\right)^{\dagger},
\end{align*}
with $U_{t,0}^{\theta,\eta}=\text{exp}[-i(g(\theta,\eta)H_{I}(Q_{i},P_{i})+f(\theta,\eta)V(Q_{i},P_{i}))t/\hbar]$
the time-evolution operator. This expression is general. For $\theta=\eta=0$,
the time-evolution operator from the standard quantum mechanics is
recovered, otherwise the evolved local thermal states will contain
noncommutative signatures.


\begin{thebibliography}{10}


\bibitem{Adesso}G. Adesso, S. Ragy, and A. R. Lee, ``Continuous variable
quantum information: Gaussian states and beyond'', Open Syst. Inf. Dyn.$\mathbf{21}$,
1440001 (2014). \url{https://doi.org/10.1142/S1230161214400010}.




\bibitem{Wang}X. Wang, T. Hiroshima, A. Tomita, and M. Hayashi,
``Quantum information with Gaussian states'', Phys. Rep. $\mathbf{448}$,
1 (2007). \url{https://doi.org/10.1016/j.physrep.2007.04.005}.

\bibitem{Adesso02}R. Nichols, Pietro Liuzzo-Scorpo, Paul A. Knott,
and Gerardo Adesso, ``Multiparameter Gaussian quantum metrology'', Phys.
Rev. A $\mathbf{98}$, 012114 (2018). \url{https://doi.org/10.1103/PhysRevA.98.012114}.

\bibitem{Yang2019}Y. Yang, ``Memory effects in quantum metrology'', Phys.
Rev. Lett. \textbf{123}, 110501 (2019). \url{https://doi.org/10.1103/PhysRevLett.123.110501}.

\bibitem{Dowling2015}J. P. Dowling and K. P. Seshadreesan, Quantum
optical technologies for metrology, sensing and imaging, J. Light.
Tech. \textbf{33}, 2359 (2015). \href{https://ieeexplore.ieee.org/document/6999929}{10.1109/JLT.2014.2386795}.


\bibitem{Chuangbook}M. A. Nielsen and I. L. Chuang, ``Quantum Computation
and Quantum Information'', (Cambridge University Press, Cambridge, 2010). \url{https://doi.org/10.1017/CBO9780511976667}.

\bibitem{Googlepaper2019}F. Arute, K. Arya, R. Babbush, et al., ``Quantum
supremacy using a programmable superconducting processor'', Nature \textbf{574},
505-510 (2019). \url{https://doi.org/10.1038/s41586-019-1666-5}.

\bibitem{Wright2019}K. Wright, K. M. Beck, S. Debnath, et al., ``Benchmarking
an 11-qubit quantum computer'', Nature Communications \textbf{10}, 5464 (2019). \url{https://doi.org/10.1038/s41467-019-13534-2}.

\bibitem{Camati2019}P. A. Camati, J. F. G. Santos, and R. M. Serra,
``Coherence effects in the performance of the quantum Otto heat engine'',
Phys. Rev. A \textbf{99}, 062103 (2019). \url{https://doi.org/10.1103/PhysRevA.99.062103}.

\bibitem{Kosloff2019}R. Dann and R. Kosloff, ``Quantum Signatures in
the Quantum Carnot Cycle'', New J. Phys. \textbf{22} 013055 (2020). \url{https://doi.org/10.1088/1367-2630/ab6876}.

\bibitem{Paternostro2019}O. Abah, M. Paternostro, ``Implications of
non-Markovian dynamics on information-driven engine'', J. Phys. Commun. \textbf{4} 085016 (2020). \url{https://doi.org/10.1088/2399-6528/abaf99}.

\bibitem{key-9}J. P. S. Peterson, T. B. Batalhão, M. Herrera, A.
M. Souza, R. S. Sarthour, I. S. Oliveira, R. M. Serra, ``Experimental
characterization of a spin quantum heat engine'', Phys. Rev. Lett. \textbf{123}, 240601 (2019). \url{https://doi.org/10.1103/PhysRevLett.123.240601}.

\bibitem{Klatzow2019}J. Klatzow, J. N. Becker, P. M. Ledingham, et
al., ``Experimental Demonstration of Quantum Effects in the Operation
of Microscopic Heat Engines'', Phys. Rev. Lett. \textbf{122}, 110601
(2019). \url{https://doi.org/10.1103/PhysRevLett.122.110601}.

\bibitem{Santos2018}J. F. G. Santos, ``Gravitational quantum well as
an effective quantum heat engine'', Eur. Phys. J. Plus \textbf{133},
321 (2018). \url{https://doi.org/10.1140/epjp/i2018-12141-8}.

\bibitem{Camati2020}P. A. Camati, J. F. G. Santos, and R. M. Serra,
``Employing Non-Markovian effects to improve the performance of a quantum
Otto refrigerator'', Phys. Rev. A \textbf{102}, 012217 (2020). \url{https://doi.org/10.1103/PhysRevA.102.012217}.

\bibitem{Reiter}M. J. Kastoryano, F. Reiter, and A. S. Sørensen,
``Dissipative Preparation of Entanglement in Optical Cavities'', Phys.
Rev. Lett. $\mathbf{106}$, 090502 (2011). \url{https://doi.org/10.1103/PhysRevLett.106.090502}.

\bibitem{Haroche2001}J. M. Raimond, M. Brune, and S. Haroche, ``Colloquium:
Manipulating quantum entanglement with atoms and photons in a cavity'',
Rev. Mod. Phys. \textbf{73}, 565 (2001). \url{https://doi.org/10.1103/RevModPhys.73.565}.

\bibitem{Walls}D. F. Walls and G. J. Milburn, ``Quantum Optics'' (Springer,
Berlin, 2008). \url{https://doi.org/10.1007/978-3-540-28574-8}.

\bibitem{Zhang2014}K. Zhang, F. Bariani, and P. Meystre, ``Theory of
an optomechanical quantum heat engine'', Phys. Rev. A \textbf{90}, 023819
(2014). \url{https://doi.org/10.1103/PhysRevA.90.023819}.

\bibitem{Kampel2013}T. P. Purdy, P.-L. Yu, R. W. Peterson, N. S.
Kampel, and C. A. Regal, ``Strong Optomechanical Squeezing of Ligh'',
Phys. Rev. X \textbf{3}, 031012 (2013). \url{https://doi.org/10.1103/PhysRevX.3.031012}.

\bibitem{Clerk2013}Ying-Dan Wang and Aashish A. Clerk, ``Reservoir-Engineered
Entanglement in Optomechanical Systems'', Phys. Rev. Lett. \textbf{110},
253601 (2013). \url{https://doi.org/10.1103/PhysRevLett.110.253601}.

\bibitem{Sage2019-1}.C. D. Bruzewicz, J. Chiaverini, R. McConnell,
and J.M. Sage, ``Trapped-Ion Quantum Computing: Progress and Challenges'',
Appl. Phys. Rev. \textbf{6}, 021314 (2019). \url{https://doi.org/10.1063/1.5088164}.

\bibitem{Clausius}R. Clausius, ``The Mechanical Theory of Heat'' (MacMillan,
London, 1879).

\bibitem{Oliveira2017}M. J. de Oliveira, ``Heat transport along a chain
of coupled quantum harmonic oscillators'', Phys. Rev. E \textbf{95},
042113 (2017). \url{https://doi.org/10.1103/PhysRevE.95.042113}.

\bibitem{Casetti2018}S. Iubini, P. Di Cintio, S. Lepri, R. Livi,
and L. Casetti, ``Heat transport in oscillator chains with long-range interactions coupled to thermal reservoirs'' Phys. Rev. E \textbf{97}, 032107 (2018). \url{https://doi.org/10.1103/PhysRevE.97.032102}.

\bibitem{Dantan2015}A. Xuereb , A. Imparato, and A. Dantan, ``Heat
transport in harmonic oscillator systems with thermal baths: application
to optomechanical arrays'', New J. Phys. \textbf{17}, 055013 (2015). \url{https://doi.org/10.1088/1367-2630/17/5/055013}.

\bibitem{Landi2019}W. T. B. Malouf, J. P. Santos, L. A. Correa, M.
Paternostro, and G. T. Landi, ``Wigner entropy production and heat transport
in linear quantum lattices'', Phys. Rev. A \textbf{99}, 052104 (2019). \url{https://doi.org/10.1103/PhysRevA.99.052104}.

\bibitem{Fiore2019}B. A. N. Akasaki, M. J. de Oliveira, and C. E.
Fiore, ``Entropy production and heat transport in harmonic chains under
time dependent periodic drivings'', Phys. Rev. E \textbf{101}, 012132  (2020). \url{https://doi.org/10.1103/PhysRevE.101.012132}.

\bibitem{Horodecki2009}R. Horodecki, P. Horodecki, M. Horodecki,
and K. Horodecki, ``Quantum entanglement'', Rev. Mod. Phys.\textbf{ 81},
865 (2009). \url{https://doi.org/10.1103/RevModPhys.81.865}.

\bibitem{Plenio2017}A. Streltsov, G. Adesso, and M. B. Plenio, ``Colloquium:
Quantum coherence as a resource'', Rev. Mod. Phys. \textbf{89}, 041003
(2017). \url{https://doi.org/10.1103/RevModPhys.89.041003}.

\bibitem{Breuer2016}H.-P. Breuer, E.-M. Laine, J. Piilo, and B. Vacchini,
``Colloquium: Non-Markovian dynamics in open quantum systems'', Rev. Mod.
Phys. \textbf{88}, 021002 (2016). \url{https://doi.org/10.1103/RevModPhys.88.021002}.

\bibitem{Modi2012}K. Modi, A. Brodutch, H. Cable, T. Paterek, and
V. Vedral, ``The classical-quantum boundary for correlations: Discord
and related measures'', Rev. Mod. Phys.\textbf{ 84}, 1655 (2012). \url{https://doi.org/10.1103/RevModPhys.84.1655}.

\bibitem{Snyder}H. S. Snyder, ``Quantized Space-Time'', Phys. Rev. \textbf{71},
38 (1946). \url{https://doi.org/10.1103/PhysRev.71.38}.

\bibitem{ref01}S. Doplicher, K. Fredenhagen and J. E. Roberts, ŽŽThe
quantum structure of spacetime at the Planck scale and quantum fields'',
Commun. Math. Phys.$\mathbf{172}$, 187 (1995). \url{https://doi.org/10.1007/BF02104515}.

\bibitem{ref02}N. Seiberg, `` Emergent Spacetime'', in The Quantum Structure of Space and Time, The - Proceedings Of The 23rd Solvay Conference in Physics (World Scientific, 2006), pp. 163?178.

\bibitem{Bernardini01}A. E. Bernardini and O. Bertolami, ``Probing
phase-space noncommutativity through quantum beating, missing information
and the thermodynamic limit'', Phys. Rev. A $\mathbf{88}$, 012101 (2013). \url{https://doi.org/10.1103/PhysRevA.88.012101}.

\bibitem{Vergara}M. Rosenbaum and J. David Vergara, ``The star-value
Equation and Wigner Distributions in Noncommutative Heisenberg algebras'',
Gen. Rel. Grav. \textbf{38}, 607 (2006). \url{https://doi.org/10.1007/s10714-006-0251-z}.

\bibitem{Orfeu01}O. Bertolami, J. G. Rosa, C. M. L. de Aragão, P.
Castorina, and D. Zappalà, ``Noncommutative Gravitational Quantum Well'',
Phys. Rev. D $\mathbf{72}$, 025010 (2005). \url{https://doi.org/10.1103/PhysRevD.72.025010}.

\bibitem{Banerjee}R. Banerjee, B. D. Roy, and S. Samanta, ``Remarks
on the Noncommutative Gravitational Quantum Well'', Phys. Rev. D74,
045015 (2006). \url{https://doi.org/10.1103/PhysRevD.74.045015}.

\bibitem{Gnatenko2019A}Kh. P. Gnatenko and V. M. Tkachuk, ``Upper bound
on the momentum scale in noncommutative phase space of canonical type'',
Eur. Phys. Lett. 127, 20008 (2019). \url{https://doi.org/10.1209/0295-5075/127/20008}.

\bibitem{Lawson2017} L. Lawson, L. Gouba, and G. Y. Avossevou, ``Two-dimensional noncommutative gravitational quantum well'', J. Phys. A: Math. Theor. \textbf{50}, 475202 (2017). \url{https://doi.org/10.1088/1751-8121/aa86c4}.

\bibitem{Leal01}P. Leal and, O. Bertolami, ``Relativistic dispersion
relation and putative metric structure in noncommutative phase-space'',
Phys. Lett. B \textbf{793}, 240 (2019). \url{https://doi.org/10.1016/j.physletb.2019.04.049}.

\bibitem{Jonas2016}J. F. G. Santos and A. E. Bernardini, ``Gaussian
fidelity distorted by external fields'', Physica A $\mathbf{445}$,
75 (2016). \url{https://doi.org/10.1016/j.physa.2015.10.033}.

\bibitem{Leal2018}O. Bertolami, A. E. Bernardini, and P. Leal, ``Quantum
information aspects of noncommutative quantum mechanics'', J. Phys.:
Conf. Ser. \textbf{952}, 012016 (2018). \url{https://doi.org/10.1088/1742-6596/952/1/012016}.

\bibitem{Leal2019}P. Leal, A. E. Bernardini, and O. Bertolami, ``Quantum
cloning and teleportation fidelity in the noncommutative phase-space'',
J. Phys. A: Math Theor. \textbf{52}, 375302 (2019). \url{https://doi.org/10.1088/1751-8121/ab359b}.

\bibitem{Paris2016}M. A. C. Rossi, T. Giani, and M. G. A. Paris,
``Probing deformed quantum commutators'', Phys. Rev. D \textbf{94}, 024014
(2016). \url{https://doi.org/10.1103/PhysRevD.94.024014}.

\bibitem{Giri2009}P. R. Giri and P. Roy, ``Non-hermitian quantum mechanics
in non-commutative space'', Eur. Phys. J. C \textbf{60}, 157 (2009). \url{https://doi.org/10.1140/epjc/s10052-009-0866-9}.

\bibitem{Santos2019A}J. F. G. dos Santos, F. S. Luiz, O. S. Duarte
and M. H. Y. Moussa, ``Non-Hermitian noncommutative quantum mechanics'',
Eur. Phys. J. Plus \textbf{134}, 332 (2019). \url{https://doi.org/10.1140/epjp/i2019-12738-3}.

\bibitem{Dey2012} S. Dey, A. Fring, and L. Gouba, ``PT-symmetric noncommutative spaces with minimal volume uncertainty relations'', J. Phys. A: Math. Theor. \textbf{45} 385302 (2012). \url{https://doi.org/10.1088/1751-8113/45/38/385302}.

\bibitem{Dey2017}S. Dey, A. Bhat, D. Momeni, M. Faizal, A. F. Ali,
T.K. Dey and A. Rehman, ``Probing noncommutative theories with quantum
optical experiments'', Nucl. Phys. B 924, 578-587 (2017). \url{https://doi.org/10.1016/j.nuclphysb.2017.09.024}.

\bibitem{Faizal2018}{]} M. Khodadi, K. Nozari, S. Dey, A. Bhat and
M. Faizal, ``A new bound on polymer quantization via an opto-mechancial
setup'', Nature Scientic Reports 8, 1659 (2018). \url{https://doi.org/10.1038/s41598-018-19181-9}.

\bibitem{Anders2016}S. Vinjanampathy and J. Anders, ``Quantum Thermodynamics'',
Cont. Phys. \textbf{57}, 545 (2016). \url{https://doi.org/10.1080/00107514.2016.1201896}.

\bibitem{Kosloff00}R. Alicki and R. Kosloff, ?Introduction to quantum thermodynamics: History and prospects,? in Thermodynamics in the Quantum Regime, edited by F. Binder et al. (Springer, Cham, 2019), pp. 1?33.

\bibitem{Deffner00}S. Deffner and S. Campbell, Quantum Thermodynamics: An Introduction to the Thermodynamics of Quantum Information (Morgan and Claypool, 2019).

\bibitem{Jonas02}J. F. G. Santos and A. E. Bernardini, ``Quantum engines
and the range of the second law of thermodynamics in the noncommutative
phase-space'', Eur. Phys. J. Plus $\mathbf{132}$, 260 (2017). \url{https://doi.org/10.1140/epjp/i2017-11538-1}.

\bibitem{Pandit2019}T. Pandit, P. Chattopadhyay, and Goutam Paul,
``Non-commutative space engine: a boost to thermodynamic processes'',\url{https://arxiv.org/abs/1911.13105}.

\bibitem{Chattopadhyay2019}P. Chattopadhyay, ``Non-Commutative space:
boon or bane for quantum engines and refrigerators'', Eur. Phys. J. Plus \textbf{135}, 302 (2020). \url{https://doi.org/10.1140/epjp/s13360-020-00318-7}.

\bibitem{Dias2009}N. C. Dias and J. N. Prata, ``Exact master equation
for a noncommutative Brownian particle'', Ann. Phys. \textbf{324} 73 (2009). \url{https://doi.org/10.1016/j.aop.2008.04.009}.

\bibitem{Almeida2017}W. O. Santos, G. M. A. Almeida, and A. M. C.
Souza, ``Noncommutative Brownian motion'', Int. J. Mod. Phys. A \textbf{32}, \url{https://doi.org/10.1142/S0217751X17501469}.
1750146 (2017).

\bibitem{Santos2019}J. F. G. Santos,``Noncommutative phase-space effects
in thermal diffusion of Gaussian states'', J. Phys. A: Math. Theor.
\textbf{52}, 405306 (2019). \url{https://doi.org/10.1088/1751-8121/ab3adb}.

\bibitem{Partovi2008}M. H. Partovi, ``Entanglement versus Stosszahlansatz:
Disappearance of the thermodynamic arrow in a high-correlation environment'',
Phys. Rev. E \textbf{77}, 021110 (2008). \url{https://doi.org/10.1103/PhysRevE.77.021110}.

\bibitem{Rudolph2012}S. Jevtic, D. Jennings, and T. Rudolph, ``Maximally
and Minimally Correlated States Attainable within a Closed Evolving
System'', Phys. Rev. Lett. \textbf{108}, 110403 (2012). \url{https://doi.org/10.1103/PhysRevLett.108.110403}.

\bibitem{Das2008}S. Das and E. C. Vagenas, ``Universality of Quantum
Gravity Corrections'', Phys. Rev. Lett \textbf{101}, 221301 (2008). \url{https://doi.org/10.1103/PhysRevLett.101.221301}.

\bibitem{Serra2019} K. Micadei, J. P. S. Peterson, A. M. Souza, et
al. ``Reversing the direction of heat flow using quantum correlations'',
Nat Commun \textbf{10}, 2456 (2019). \url{https://doi.org/10.1038/s41467-019-10333-7}.

\bibitem{Conti2019}G. Marcucci and C. Conti, ``Simulating general relativity
and non-commutative geometry by nonparaxial quantumfluids'', New. J.
Phys. \textbf{21}, 123038 (2019). \url{https://doi.org/10.1088/1367-2630/ab5da8}.

\bibitem{Rojas2001}J. Gamboa, M. Loewe, and J. C. Rojas, ``Noncommutative
quantum mechanics'', Phys. Rev. D \textbf{64}, 067901 (2001). \url{https://doi.org/10.1103/PhysRevD.64.067901}.

\bibitem{Bastos2008}C. Bastos, O. Bertolami, N. C. Dias, J. N. Prata,
``Weyl-Wigner Formulation of Noncommutative Quantum Mechanics'', J. Math.
Phys. \textbf{49}, 072101 (2008). \url{https://doi.org/10.1063/1.2944996}.

\bibitem{Gouba2010}C. M. Rohwer, K. G. Zloshchastiev, L. Gouba, and
F. G. Scholtz, ``Noncommutative quantum mechanics---a perspective on
structure and spatial extent'', J. Phys. A: Math. Theor. \textbf{43},
345302 (2010). \url{https://doi.org/10.1088/1751-8113/43/34/345302}.

\bibitem{Gouba2016}L. Gouba, ``A comparative review of four formulations
of noncommutative quantum mechanics'', Int. J. Mod. Phys. \textbf{31},
1630025 (2016). \url{https://doi.org/10.1142/S0217751X16300258}.

\bibitem{Saha}A. Saha, S. Gangopadhyay, and S. Saha, ``Noncommutative
quantum mechanics of a harmonic oscillator under linearized gravitational
waves'', Phys. Rev. D $\mathbf{83}$, 025004 (2011). \url{https://doi.org/10.1103/PhysRevD.83.025004}.

\bibitem{Bastos}C. Bastos, N.C. Dias, and J.N. Prata, ``Wigner Measures
in Noncommutative Quantum Mechanics'', Commun. Math. Phys. $\mathbf{299}$,
3 (2010). \url{https://doi.org/10.1007/s00220-010-1109-5}.

\bibitem{Bastos02}C. Bastos, O. Bertolami, and N. Dias, J. Prata,
``Noncommutative Graphene'', Int. J. Mod. Phys. A $\mathbf{28}$, 16 (2013). \url{https://doi.org/10.1142/S0217751X13500644}.

\bibitem{Andreas}J. B. Geloun, F. G. Scholtz, ``Coherent states in
noncommutative quantum mechanics'', J. Math. Phys. $\mathbf{50}$, 043505
(2009). \url{https://doi.org/10.1063/1.3105926}.

\bibitem{Liang2019}T. Harko and S.-D. Liang, ``Energy-dependent noncommutative
quantum mechanics'', Eur. Phys. J. C \textbf{79}, 300 (2019). \url{https://doi.org/10.1140/epjc/s10052-019-6794-4}.

\bibitem{Bastos2008-1}C. Bastos, O. Bertolami, N. C. Dias, and J.
N. Prata, ``Phase-Space Noncommutative Quantum Cosmology'', Phys. Rev.
D \textbf{78}, 023516 (2008). \url{https://doi.org/10.1103/PhysRevD.78.023516}.

\bibitem{Serafini}A. Serafini, ``Quantum Continuous Variables. A primer
of Theoretical Methods'', (CRC Press, Boca Raton, 2017). \url{https://doi.org/10.1201/9781315118727}.

\bibitem{Orfeu02}C. Bastos and O.Bertolami, ``Berry phase in the gravitational
quantum well and the Seiberg--Witten map'', Phys. Lett. A $\mathbf{372}$,
34 (2008). \url{https://doi.org/10.1016/j.physleta.2008.06.073}.

\bibitem{Santos2015}J. F. G. Santos, A. E. Bernardini, and C. Bastos,
``Probing phase-space noncommutativity through quantum mechanics and
thermodynamics of free particles and quantum rotors'', Physica A $\mathbf{438}$,
340 (2015). \url{https://doi.org/10.1016/j.physa.2015.07.009}.

\bibitem{Gamboa}J. Gamboa, M. Loewe, and J. C. Rojas, ``Noncommutative quantum mechanics'', Phys. Rev.
D $\mathbf{64}$, 067901 (2001). \url{https://doi.org/10.1103/PhysRevD.64.067901}.

\bibitem{Jacak1998}L. Jacak, P. Hawrylak, and A. Wjs, ``Quantum Dots''
(Springer-Verlag, 1998). \href{https://www.springer.com/gp/book/9783642720048#aboutBook}{10.1007/978-3-642-72002-4}.

\bibitem{Simon}R. Simon, E. E. G. Sudarsan, and N. Makunda, ``Gaussian-Wigner
distributions in quantum mechanics and optics'', Phys. Rev. A $\mathbf{36}$,
3868 (1987). \url{https://doi.org/10.1103/PhysRevA.56.5042}.
\end{thebibliography}
\end{document}